# Mixer Hamiltonian with QAOA for Max *k*-coloring: numerical evaluations

E. Bourreau[1], G. Fleury[2], P. Lacomme[2]

[2]*LIRMM, Université de Montpellier, CNRS*
*161 rue Ada, 34392 Montpellier Cedex 5*
Eric.Bourreau@lirmm.fr

[2]*Université Clermont-Auvergne, Clermont-Auvergne-INP, LIMOS (UMR CNRS 6158),*
*1 rue de la Chebarde,*
*63178 Aubière Cedex, France*
gerard.fleury@isima.fr, philippe.lacomme@isima.fr

**Abstract.** This paper concerns quantum heuristics based on Mixer Hamiltonians that allow to restrict investigation on a specific subspace. Mixer Hamiltonian based approaches can be included in QAOA algorithm and we can state that Mixer Hamiltonians are mapping functions from the set of qubit-strings to the set of solutions. Mixer Hamiltonian offers an approach very similar to indirect representations commonly used in routing or in scheduling community for decades. After the initial publication of Cheng et al. in 1996 (Cheng et al., 1996), numerous propositions in OR lies on 1-to-$n$ mapping functions, including the split algorithm that transform one TSP solution into a VRP solution.

The objective is at first to give a compact and readable presentation of these Mixer Hamiltonians considering the functional analogies that exist between the OR community practices and the quantum field. Our experiments encompass numerical evaluations of circuit using the Qiskit library of IBM meeting the theoretical considerations.

## 1. Introduction

Quantum optimization is a new and attractive area with potential significant implication in operation research. Minimization problems can now be investigated using quantum metaheuristics with the promise of a strongly effective approach avoiding trapping into local minima that standard local search. Simulated Annealing based methods efficiency (commonly used in OR field) is performed by slowly reducing to 0 a parameter $t$ that permits to follow the potential barriers. The metaheuristic methods differ by the way used to avoid premature convergence to local minima and to conserve a strong capacity in search space investigation. The sequel simulated annealing is an example of a large class of metaheuristics that encompasses (but are not limited to): memetic algorithms, GRASP, VNS…

From a quantum mechanics point of view, quantum fluctuations are similar to thermal fluctuation. Quantum mechanics differ from classical approaches considering that waves can tunnel through potential barriers (energy) (Martoòák et al., 2004). In the last few years several quantum metaheuristics have been introduced coming from the quantum physic community, defining a family of quantum approximate algorithms that encompasses for example the sequel Adiabatic based Algorithms that provides an approximate solution of the Schrödinger equation.

Lately (Farhi et al., 2014) have introduced a new class of algorithms based on alternation between two families of operators referred to as Hamiltonian and mixing Hamiltonian leading to Quantum Approximate

Optimization Algorithms referred to as QAOA. Such algorithms are hybrid algorithms where the search space investigation is devoted to a classical computer to optimize a set of parameters and the distribution of probabilities evaluation is achieved by a quantum device. This algorithm that does no support local search consideration and that provides a full search space investigation, has been extended in the famed publication of (Farhi and Harrow, 2019). They defined new ansatz that support the exploration of the feasible subspace only, meaning that hard constraints are satisfied by definition. This approach is very similar to classical approaches of the OR community since the exploration of the feasible subspace is achieved by a careful definition of classical operation including for example permutation of qubit into the qubit-string used to model solutions;

## 2. QAOA based approach

Quantum Approximate Optimization Algorithms (Farhi et al, 2014) take advantage of alternations between the cost function investigation which is modeled by a Hamiltonian $H_P$ from one side and a driver Hamiltonian operator $H_D$. The Quantum Alternating Operator Ansatz (Hadfield et al, 2018) takes into consideration a general parameterized family of unitary operators to efficiently modelize the Hamiltonian. This algorithm supports a restricted number of states and create an efficient alternative to the Adiabatic Optimization

### 2.1. QAOA Principles

QAOA based approaches can be seen as metaheuristic-based methods taking advantages of time-discretization of adiabatic computing. It is a hybrid algorithm because it combines:

- a Quantum parametrized circuit executed on a quantum machine,
- a Classical meta-optimization loop executed on a classical machine.

As introduced by (Schrödinger, 1926) the wave function evolution of a quantum-mechanical system is given by

$$\frac{\partial}{\partial t}|\psi(x,t)\rangle = -\frac{i}{\hbar}.H(t).|\psi(x,t)\rangle$$

where the energy is defined by $H(t)$, $\hbar$ is derived from Plank constant and $|\psi(x,t)\rangle$ are states vectors. If $H$ is time independent the solution is $|\psi_t\rangle = e^{-\frac{i}{\hbar}.t.H}.|\psi_0\rangle$. Note that the solution is $|\psi_T\rangle = e^{-\frac{i}{\hbar}.\int_0^T H(u).du}.|\psi_0\rangle$ in the general time dependent situation. Describing a problem with a Hamiltonian $H$ and an initial state $|\psi_0\rangle$ allows to compute the ground state.

A specific resolution can be achieved considering an iterative approximation of the Schrödinger solution taking advantages of a very specific $H$ Hamiltonian (based on $Z_i, Z_iZ_j$ operators). In Adiabatic Quantum Optimization, the system must be tuned in a ground state of one Hamiltonian $H_D$ commonly referred to as the "driver Hamiltonian" where this Hamiltonian should not commute with the Hamiltonian that model the combinatorial problem. It requires to slowly tuned from $H_D$ to $H_P$ using an interpolation based on one parameter $s(t)$ assumed to smoothly decrease from 1 to 0: $H(t) = s(t).H_D + [1 - s(t)].H_P$ (Farih et al., 2000).

Adiabatic quantum optimization and all quantum optimization based annealing approaches are limited to small-scale instances due to the slow evolution that is required from $H_D$ to $H_P$. In 2014 (Farih et al., 2014) have introduced a new trend of approaches referred to as quantum heuristics leading to a more compact decomposition scheme denoted Quantum Approximate Optimization Algorithm (QAOA). QAOA received a considerable amount of attention including but not limited to (Farih and Harrow, 2019), (Yang and al., 2017), (Jiang et al., 2017), (Wecker et al., 2016) and (Wang et al., 2018).

Lately (Farhi and Harrow, 2019) prove that the output distribution obtained at the end of one QAOA algorithm cannot be efficiently approximated using one classical algorithm because of the undue complexity of the required algorithms. Over the past few years QAOA has attracted attention to many researchers, because it seems to be one candidate for demonstration of quantum supremacy as stressed by (Farhi and Harrow, 2019). A quantum supremacy demonstration consists in defining one circuit with results requiring an exponential number of operations into a classical algorithm.

## 2.2. Modelling function to Hamiltonian

QAOA seeks to solve a hard optimization problem i.e. minimizing or maximizing one objective function $f(x)$ that is assumed to act on $n-bits$ strings $x$. QAOA is based on $p$ consecutive iterations of one Hamiltonian $H_P$ cumulated with a driver Hamiltonian $H_D$, where this weighted sum of Hamiltonian terms varies in time. The Hamiltonian $H$ maps the function $f$ with $2^n$ eigenvalues that model the $2^n$ values of $f$. The optimal solution (i.e. the extremal value of $f$) is an eigenvalue of $H$ and $H$ is satisfying:
$$H.|\text{x}\rangle = f(x)|\text{x}\rangle$$
Because a Hamiltonian is a Hermitian operator it has a spectral decomposition: $H = \sum_i e_i |e_i\rangle\langle e_i|$ where $|e_i\rangle$ is the $i^{th}$ basis vector.

The Hamiltonian is defined with Pauli operator-basis and takes advantages of the Pauli $Z$ leading to expression
$$H = \alpha_0.I + \alpha_1.Z_0.Z_1 + \alpha_2.Z_0.Z_2 \ldots$$
where $\alpha_i$ are real numbers.

Note that Hadfield in 2021 (Hadfield, 2021) gives a concise description of rules for composing Hamiltonians representing clauses that model clauses or functions including Boolean formulas. A Hamiltonian is implemented into a quantum circuit by deriving $U_H(t) = e^{-i.H.t}$ with $t \in [0; 2\pi]$ and using both CNOT and Z-rotations. $t$ refers to the weight in the iterative search process of QAOA.

## 2.3. Search space investigation

The $\vec{\beta}$ and $\vec{\gamma}$ weights parametrized a quantum state $|\varphi(\vec{\beta},\vec{\gamma})\rangle$ that defines a solution $y$ with probability $|\langle y||\varphi(\vec{\beta},\vec{\gamma})\rangle|^2$ and an expectation value $\langle \varphi(\vec{\beta},\vec{\gamma})|H|\varphi(\vec{\beta},\vec{\gamma})\rangle$ estimated by sampling. This value estimate the average cost of the problem $P : C^p\left(\overrightarrow{\beta},\overrightarrow{\gamma}\right)$. The overall algorithm description is illustrated in Figure 1.

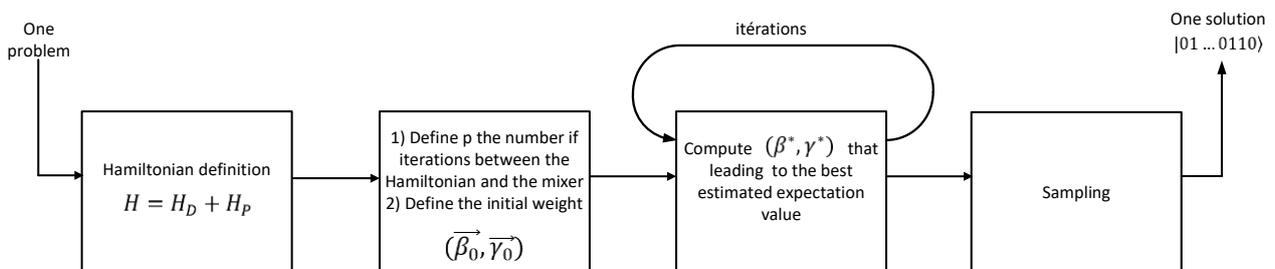

Fig. 1. QAOA principles

We encompass that QAOA efficiency strongly relies on some key-points (figure 2):

- providing a good ratio between the estimation of quality of $C^p\left(\vec{\beta},\vec{\gamma}\right)$ versus the number of shots is required and must be tuned carefully.
- The last computed distribution $|\psi(\vec{\beta^*},\vec{\gamma^*})\rangle$ must be collected on a subset of solutions strongly smaller than the total number of solutions to avoid a costly inefficient enumerations i.e. the algorithm has converged to the optimal solutions and quasi solutions.

In the following Figure 2 we illustrate two separated search spaces computed in a hybrid way on two separated machines (one classical, one quantum computer).

First, we have to compute analytically a Hamiltonian $H$ with 2 angles $(\vec{\beta},\vec{\gamma})$. Second, we should be capable to convert this as a circuit. Finally we have to estimate by sampling the expected value $C^p\left(\vec{\beta},\vec{\gamma}\right)$. This give information to investigate new value parameter to compute a new Hamiltonian.

The success of QAOA relies on the algorithm capacity to find a good set of $2.p$ parameters and by consequence an efficient search strategy is required in this variational phase.

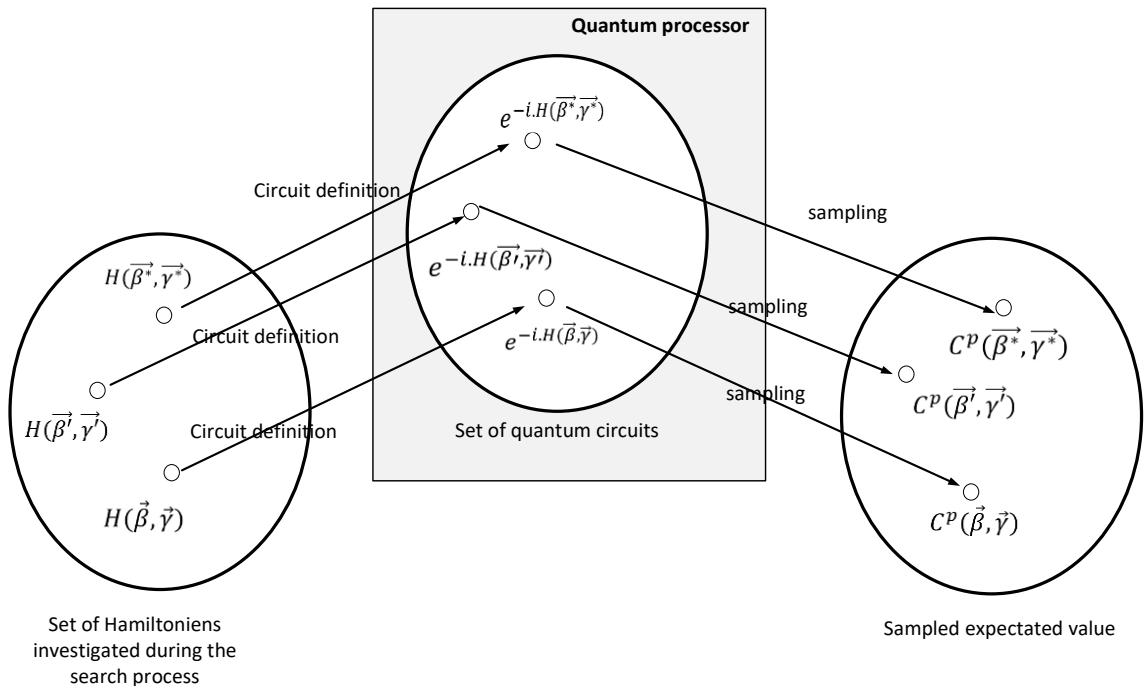

Fig. 2. Links between the weighted Hamiltonian and the expected cost.

## 3. Mixer Hamiltonian with QAOA

A wide majority of Operational Research problems can be defined to both minimization of one objective function and satisfaction of problem dependent constraints. Resolution approaches are based on exact methods, heuristics or metaheuristics depending on the problem and the computation time available to provide a solution. For a practical point of view, heuristics and metaheuristics are commonly used in resolution of large instances and they address the following key-points during the search space investigation:

- generation of good initial solutions by powerful constructive heuristics dedicated to the problem;
- application of a local search to solutions to find local minima and by consequence favoring convergence;

- diversification mechanisms to avoid premature convergence and search space trap;
- Indirect representation of solutions to relax constraints and alternate between several search spaces.

In the next section, we will focus on representations of solutions.

*3.1. Mapping function in O.R. field*

Quantum annealing optimization principles (QA) consists into a smooth decreases of quantum fluctuations to traverse barriers in the energy landscapes investigating the global minima of the function.

Quantum annealers are limited in resolution of optimization problems by the Hamiltonian circuit size and specifically by constraints included in the objective function using penalty terms. Driven Hamiltonian based approaches are designed to avoid this penalty term and the all-to-all connectivity and to defined new Hamiltonian that commutes with the constraints.

Mixer Hamiltonians are tuned to favor investigation of operational researches where the feasible subspace is strongly lower that the full space (set of $n$-bit strings used for $H$) using a mapping from initial states to final solutions. Investigating only feasible subspace has received attention to the Operation Research (O.R.) community for decades and such approaches have been successful in several research areas leading to very efficient metaheuristics.

In Job-Shop Scheduling a common indirect representation is based on the Bierwith vector (Bierwith, 1996) that can be transform (in $O(n)$) into one oriented disjunctive graph that model one solution of the job-shop. We can also cite scheduling problems based on disjunctive graph, the local search approaches take advantages of the longest path (see the block definition of Grabowski for example) (Grabowski et al., 1986). In flow based scheduling problems (RCPSP for example) the local search is based on the cut (Artigues et al., 2003), and in routing problems based on the geometry of solutions. The diversification mechanisms are metaheuristic dependent, for example they are based on the temperature for the simulated annealing (SA) or mutation in memetic algorithms (MA) (Moscato, 1999), and permits to accept transitions from one solution to worst solutions with a non-zero probability to cross the barriers in the objective function landscape.

In VRP a well-known representation is the giant trip that can be transformed into one VRP solution using the Split Algorithm (Lacomme et al., 2001). The Split algorithm defines a mapping from the set of giant trip of a TSP (Traveling Salesman Problem) to solutions of the VRP and the metaheuristic based approaches manipulate efficiently the set of giant trips only.
Cheng et al. in 1996 (Cheng et al., 1996) are the very first who made a full analysis of non-string coding approach and decoding mechanisms in the global context of constraint optimization.

Their remarks encompass both scheduling and routing by emphasizing the principle of a mapping perspective from the problem constraint in order to have indirect representation that generate either an initial phase and evolutionary process production of feasible solution only.

The split algorithms (see (Prins et al., 2014)) are one classical approach in routing by transforming a giant trip of the TSP into (for example) a solution of the VRP.

In this context most order-first/split-second methods generate in a first phase an indirect solution representation, referred to as giant tour or task ordering. Feasible routes are deduced from this indirect representation in a second phase. This approach offers the following advantages: (i) any solution of the routing problem has an indirect string representation (ISR); (ii) using a splitting procedure, each ISR can be decoded into a solution to the original problem, and this splitting can be made optimally and (iii) there exists at least one "optimal" ISR, i.e., one that gives an optimal solution to the original problem after splitting. It has been proved by (Prins, 2004) that the local search operator must be defined both in modification of a VRP solution including basic moves (2-OPT, insert….) and also moves applied to the giant trip. The global performance of the local search is based on alternation between the two spaces.

The Mixer Hamiltonians are on the same trend of research considering that a careful definition of the mapping function, should permit to transform one qubit-strings into one solution to favor exploration of the "feasible" search space only. Contrary to the OR classical field, the explicit local search dynamic is embedded by the Mixer Hamiltonien.

*3.2. Mixer Hamiltonian: Mapping function to feasible solutions*

A wide and important class of optimization problems seeks to minimize an objective function $f: F \to \mathbb{R}$ to be optimized under a set of constraints.

$$f(x) = \sum_{i=1}^{n} c_i . x_i$$
$$\forall j = 1..m \; g_j(x) \geq b_j$$

Despite the fact that it is possible to create a QAOA circuit for an objective function of order higher that quadratic, and for a large majority of classical O.R. quadratic expressions can be used:

$$f(x) = \sum_{i=1}^{n} c_i . x_i + \sum_{j=1}^{m} A_j . \left(b_j - g_j(x)\right)^2$$

Using a classical Hamiltonian construction, an initial state $|s\rangle = |+\rangle^{\otimes n}$ gives an initial Hamiltonian decomposition composed only of $I$ and $X$ Pauli gates applied to $H$ to $|0\rangle^{\otimes n}$ leading to one uniform distribution over the $2^n$ values of $f$.

The major drawback of the classical Hamiltonian formulation concerns the weighted quadratic terms in the objective function meaning that the search space investigation is not limited to "feasible" solutions but also encompasses $x_i$ assignments where the constraints do not hold. The second drawback concerns theweighted parameters $A_j$ that must be large enough to favor convergence to the feasible solution set, but must permit to avoid to be trapped into local minima.

The Mixer Hamiltonians extend the initial proposal of (Hen and Spedalieri, 2016) in the adiabatic context where they identify the $H_{jk} = X_j . X_k + Y_j . Y_k$ permitting to restrict the state evolution to feasible subspace for some problems. Note that $H_{jk}$ and $H_{ik}$ do not commute in general and that trotterization is required using the Masuo's proposition (Suzuki, 1976). A more compact and readable demonstration has been introduced lately by Barthel et Zhang in 2019 (Barthel and Zhang, 2019) and it concludes by the following proposition.

**Proposition.** $\lim_{n \to \infty} \left(e^{i.A.t/n} . e^{i.B.t/n}\right)^n = e^{i.(A+B).t}$
And as a consequence $e^{i.(A+B).t} \approx e^{i.(A).t} . e^{i.(B).t}$
where $A$ and $B$ are matrices.

Defining one Mixer Hamiltonian consists in including restriction in search space investigation in the mixing operator, is supposed to reduce the efficiency required in terms of gates and avoiding inclusion of extra weighted terms in the objective.

*3.3. XY-Hamiltonian analysis*

The *XY*-Hamiltonian

$$H(\gamma) = \sum_{i=1}^{i=N}[(1+\gamma).X_i.X_{i+1} + (1-\gamma).Y_i.Y_{i+1}]$$

can be expressed as a quadratic form in creation and annihilation operators and can be diagonalized to full describe the complete set of states, excitation energies.

It is very similar to the generalized Heisenberg model described by the Hamiltonian:

$$H(\gamma) = \sum_{i=1}^{i=N}[(1+\gamma).X_i.X_{i+1} + (1-\gamma).Y_i.Y_{i+1} + (1-\gamma).Z_i.Z_{i+1}]$$

Note that when $\gamma \to 1$, this Hamiltonian tends to the Ising model where the *x*-componants are fully ordered. The transverse terms (i.e. $Y_i.Y_{i+1}$ and $Z_i.Z_{i+1}$) consist in favoring the ordered of both *y*-components and *z*-components. A complete description of this operator has been introduced in antiferromagnetic chain by (Lieb and al., 1961).

*Qudit operators and basic principles*

Let us denote $|\varphi\rangle = |\varphi_1\varphi_2\varphi_3\rangle$ a qubit composed of 3 qubits $|\varphi_1\rangle = \begin{pmatrix}a\\b\end{pmatrix}$, $|\varphi_2\rangle = \begin{pmatrix}c\\d\end{pmatrix}$, $|\varphi_3\rangle = \begin{pmatrix}e\\f\end{pmatrix}$. The operator $X_1X_2 + Y_1Y_2$ is the application of operator $X_1 + Y_1$ to qubit 1 and $X_2 + Y_2$ to qubit 2.

$X + Y$ can be rewritten:

$$\frac{1}{\sqrt{2}}(X+Y) = \frac{1}{\sqrt{2}}\left[\begin{pmatrix}0 & 1\\1 & 0\end{pmatrix} + \begin{pmatrix}0 & -i\\+i & 0\end{pmatrix}\right] = \begin{pmatrix}0 & \frac{1}{\sqrt{2}}(1-i)\\ \frac{1}{\sqrt{2}}(1+i) & 0\end{pmatrix}$$

For our qudit $|\varphi\rangle = \begin{pmatrix}a\\b\\c\\d\\e\\f\end{pmatrix}$, the operator $X_1X_2 + Y_1Y_2$ transforms $|\varphi\rangle$ into a new state

$$\begin{pmatrix} \frac{1}{\sqrt{2}}(1-i).b \\ \frac{1}{\sqrt{2}}(1+i).a \\ \frac{1}{\sqrt{2}}(1-i).d \\ \frac{1}{\sqrt{2}}(1+i).c \\ e \\ f \end{pmatrix}$$

If we apply the operator $X_2 X_3 + Y_2 Y_3$, $|\varphi\rangle$ is now

$$\begin{pmatrix} \frac{1}{\sqrt{2}}(1-i).b \\ \frac{1}{\sqrt{2}}(1+i).a \\ c \\ d \\ \frac{1}{\sqrt{2}}(1-i).f \\ \frac{1}{\sqrt{2}}(1+i).e \end{pmatrix}$$

If we successively apply $X_i X_{i+1} + Y_i Y_{i+1}$ to a basis qubit, this will allow to superpose the current state with a new one where the 1 value is successively switched to the next position in the qubit.

To implement this operator, we will manipulate:
$$e^{-i.t.\gamma.X_1.X_2} = \cos(t.\gamma).Id - i.\sin(t.\gamma).X_1.X_2$$

$$e^{-i.t.\gamma.X_1.X_2} = \cos(t.\gamma).\begin{pmatrix} 1 & 0 & 0 & 0 \\ 0 & 1 & 0 & 0 \\ 0 & 0 & 1 & 0 \\ 0 & 0 & 0 & 1 \end{pmatrix} - i.\sin(t.\gamma).\begin{pmatrix} 0 & 0 & 0 & 1 \\ 0 & 0 & 1 & 0 \\ 0 & 1 & 0 & 0 \\ 1 & 0 & 0 & 0 \end{pmatrix}$$

$$e^{-i.t.\gamma.X_1.X_2} = \begin{pmatrix} \cos(t.\gamma) & 0 & 0 & -i.\sin(t.\gamma) \\ 0 & \cos(t.\gamma) & -i.\sin(t.\gamma) & 0 \\ 0 & -i.\sin(t.\gamma) & \cos(t.\gamma) & 0 \\ -i.\sin(t.\gamma) & 0 & 0 & \cos(t.\gamma) \end{pmatrix}$$

This is easily converted to the following tensor product, using quantum gates CNOT ($CX_{ij}$), Hadamard ($H$) and Rotation on $X$ or on $Z$ axis:

$$H \otimes H.CX_{1,2}.R_Z^2(t.2.\gamma).CX_{1,2}.H \otimes H \text{ for } X_1 X_2 \text{ and}$$
$$R_X^1\left(\frac{\pi}{2}\right) \otimes R_X^2\left(\frac{\pi}{2}\right).CX_{1,2}.R_Z^2(t.2.\gamma).CX_{1,2}.R_X^1\left(-\frac{\pi}{2}\right) \otimes R_X^2\left(-\frac{\pi}{2}\right) \text{ for } Y_1 Y_2.$$

4. **Graph coloring and max k-coloring**

*4.1 Definition*
Coloring problem want to find the minimum value of the chromatic number $\chi(G)$ where there exists a $\chi(G)$ coloring for a graph $G$. More formally, for a graph $G(V, E)$, a proper coloring $\varphi: V(G) \to C$ exists with $\varphi(i) \neq \varphi(j)$ if $(i,j) \in E(G)$. If there exists $(i,j) \in E(G)$, with $\varphi(i) = \varphi(j)$, then $\varphi$ is an improper coloring of $G$. Figure 3 gives a proper coloring solution for a 5 nodes graph.

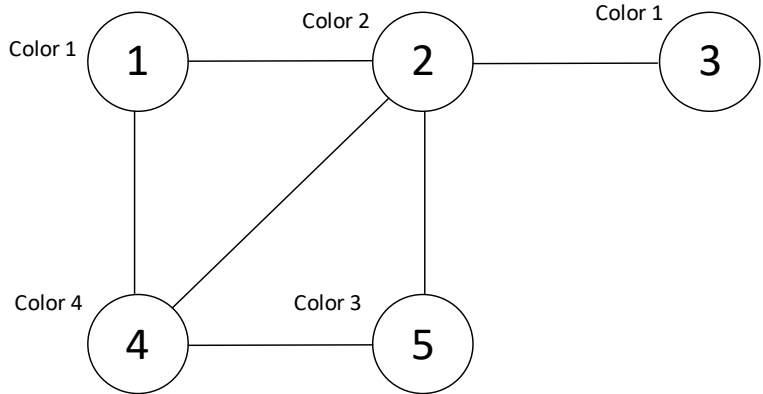

Fig. 3. Example of proper coloring

The graph coloring is the corner stone of many OR problems including the timetabling, scheduling and it has received a considerable amount of attention for its utility.

The classical formulation relies on binary variables with $x_{ik} = 1$ if the color $k$ is assigned to node $i$. With such a representation, the solution of figure 5 can be modelized by the following bit-string
$$1000 - 0100 - 1000 - 0001 - 0010$$
Each group of 4 digits is a Dirac representation of the color number k, each 5 group of digits represent a vertex. Here $c_1 = 1000$, $c_2 = 0100$, $c_3 = 1000$, $c_4 = 0001$ and $c_5 = 0010$.

The objective of a max K-coloring dealing with proper and improper colorings is to maximize the number of adjacent nodes having different color between 1 to K

$$Max \sum_{(i,j)\in E} \sum_{k=1}^{K} (x_{ik} \neq x_{ik})$$

Such that

C1. $$\forall i = 1..n, \sum_{k=1}^{K} x_{ik} = 1$$

The problem can be rewritten into a minimization of a quadratic formulation by considering:
- $S_k = \sum_{(i,j)\in E}^{n} x_{ik}.x_{jk}$ as the number of adjacent nodes with the same color
- the weighted penalized term, $A.\sum_{i=1}^{n}(1 - \sum_{k=1}^{K} x_{ik})^2$

$$S_k = \sum_{(i,j)\in E}^{n} x_{ik}.x_{jk} = 0$$

**Quadratic formulation:**

$$Min \sum_{k=1}^{K} \sum_{(i,j)\in E}^{n} x_{ik}.x_{jk} + A.\sum_{i=1}^{n}\left(1 - \sum_{k=1}^{K} x_{ik}\right)^2$$

*4.2. Hamiltonian definition*

The mapping to the binary variables is obtained considering $H = \frac{1}{2}Id - \frac{1}{2}Z$ which has two eigenvalues 0 and 1 that model $|0\rangle$ and $H = \frac{1}{2}Id + \frac{1}{2}Z$ which has the eigenvalue 1 and 0, that model $|1\rangle$. Reformulation of the function of our coloring problem with $Z_j$ variables leads to:

$$H_P = \sum_{k=1}^{K} \sum_{(i,j) \in E}^{n} \frac{1}{4}\left(Id - Z_{jk} - Z_{ik} + Z_{ik}.Z_{jk}\right) + A.\sum_{i=1}^{n}\left(1 - \sum_{k=1}^{K}\frac{1}{2}(Id - Z_{ik})\right)^2$$

The driver operator

$$H_D = -\sum_{i=1}^{n}\sum_{k=1}^{K} X_{ik}$$

is applied to $|\psi_0\rangle = H.|0\rangle^{\otimes n}$ to define the initial fundamental ground state.

The qubits-string encoding a solution is $n \times K$ qubit long and the Hamiltonian mapping function encodes a total search space of $2^{n \times K}$ colorings of $G$ that encompasses both proper and improper colorings. For an operational research point of view, the mapping function is not an indirect coding function and it is strongly different than the Split procedure of the VRP or than the efficient Bierwith vector encoding for the Job-Shop.

The previous remarks push us to consider that the Driven Hamiltonian must have afforded restrictions of exploration to coloring with only one color per node, so that the promise to only switch over coloring that satisfies the constraint C1, enabling us to control which qubit-strings must be iteratively investigated.

*4.2. A Mixer Hamiltonian definition*

The mixing operator

$$H_D = -\sum_{i=1}^{n}\sum_{j=1}^{K}\left(X_{ij}X_{ij+1} + Y_{ij}Y_{ij+1}\right)$$

permits to have Driven Hamiltonian that is a 1-to-1 mapping function i.e. an operator that maps one qubit-string on the desired subspace where the C1 constraint holds (figure 4).

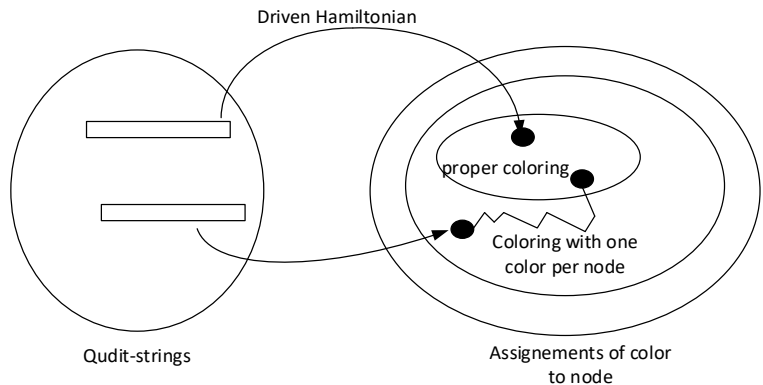

Fig. 4. Example of proper coloring

The next sections are dedicated to numerical experiments achieved on the simulator provided by IBM using the Qiskit library. The objective is to evaluate and to illustrate the efficiency for a Mixer Hamiltonian on few instances of graph coloring.

## 5. Experimental validation

With the following widget examples, the goal is to illustrate the bonus achieve by a Mixer Hamiltonian but it is not a performance analysis of quantum v.s. classical optimization since only tiny examples are tractable on quantum simulator. Of course, any classical metaheuristic will overfit the QAOA results. In a near future, if scaling in the number of (denoised) qubits and long enough coherent qubits time occurs, this evaluation will give insight on the way to model efficiency combinatorial optimization problems.

### 5.1. Numerical experiments with QAOA with 5 nodes / 6 edges and 4 colors
First experiment is achieved on the graph of figure 5 with the following set of parameters:
- $p = 3$ (iterations of Hamiltonian and mixer Hamiltonian);
- $A = 100$ is the weight assigned to the quadratic term in the objective function (it can be lower by this value improve the readability of the numeric solutions) ;
- $\overrightarrow{\beta_0} = (0.0, \ldots 0.0)$ and $\overrightarrow{\gamma_0} = (0.0, \ldots 0.0)$ ;
- Cobyla is the iterative numerical method used to minimize the estimated expectation value with 500 iterations;
- 100 samplings are used to estimate $C^p\left(\overrightarrow{\beta}, \overrightarrow{\gamma}\right)$ at step $i$;
- 100 samplings of $|\psi(\overrightarrow{\beta^*}, \overrightarrow{\gamma^*})\rangle$ are achieved at the end of QAOA to compute $|s^*\rangle$.

Results are display in table 1. A very low number of samplings at the end of the QAOA algorithm collects subset of solutions strongly smaller than the total number of solutions and avoid a costly inefficient total enumeration. The very small value 100 used permits to have the optimal solution and 100 samplings of $|\psi(\overrightarrow{\beta^*}, \overrightarrow{\gamma^*})\rangle$ samplings is sufficient in this very specific run. Note that with very low number of samplings significant differences could appear in the estimation of the distribution and that the optimal solution could not appear.

**Table 1**

One possible run with $A = 100$ and 100 samplings of $|\psi(\overrightarrow{\beta^*}, \overrightarrow{\gamma^*})\rangle$

| Solution cost | Probability |
|---:|---:|
| 0 | 1 |
| 101 | 1 |
| 102 | 3 |
| 103 | 3 |
| 201 | 7 |
| 202 | 4 |
| 203 | 4 |
| 204 | 1 |
| 300 | 1 |
| 301 | 1 |
| … | … |

The overall trend of the probability distribution can be estimated with larger number of samplings and conclusively push into considering that the emphasis of probabilities has a global trend to the improve probability of solutions where the constraint C1 holds. Note that it pushes us into considering that QAOA has aggregated probability on solutions closed to the optimal ones. The experiment of figure 7 has been achieved with same parameters and 1000 samplings of $|\psi(\vec{\beta^*},\vec{\gamma^*})\rangle$

**Table 2**
Sampling of $|\psi(\vec{\beta^*},\vec{\gamma^*})\rangle$ with 1000 sampling before and after QAOA process

| Solution cost | Probability at the end of QAOA | Cumulative probabilities at the end of QAOA | Probability at the beginning of QAOA | Cumulative probabilities at the end of QAOA |
|---|---|---|---|---|
| 0 | 0.2 | 0.0 | 0.0 | 0.0 |
| 1 | 0.3 | 0.5 | 0.1 | 0.1 |
| 2 | 0.1 | 0.6 | 0.0 | 0.1 |
| 100 | 0.3 | 0.9 | 0.0 | 0.1 |
| 101 | 1.6 | 2.5 | 0.0 | 0.1 |
| 102 | 1.3 | 3.8 | 0.2 | 0.3 |
| 103 | 0.9 | 4.7 | 0.0 | 0.3 |
| 104 | 0.2 | 4.9 | 0.0 | 0.3 |
| 106 | 0.1 | 5 | 0.0 | 0.3 |
| 200 | 1.3 | 6.3 | 0.3 | 0.6 |
| … | … | | | |

To identify the combinatoric problems more closely and the consequence with the objective to propose the adequate solutions, we may consider a graph $G$ with 5 nodes and 4 colors for which the total number of qubit-strings is $2^{20} = 1\,048\,576$. For one node, there is only 4 assignments of exactly one color, over the $2^4 = 16$ possible assignments. The probability to have one color only assigned to each node is about $\left(\frac{4}{16}\right)^5 = 0.00098\ i.e.\ 0.098\%$.

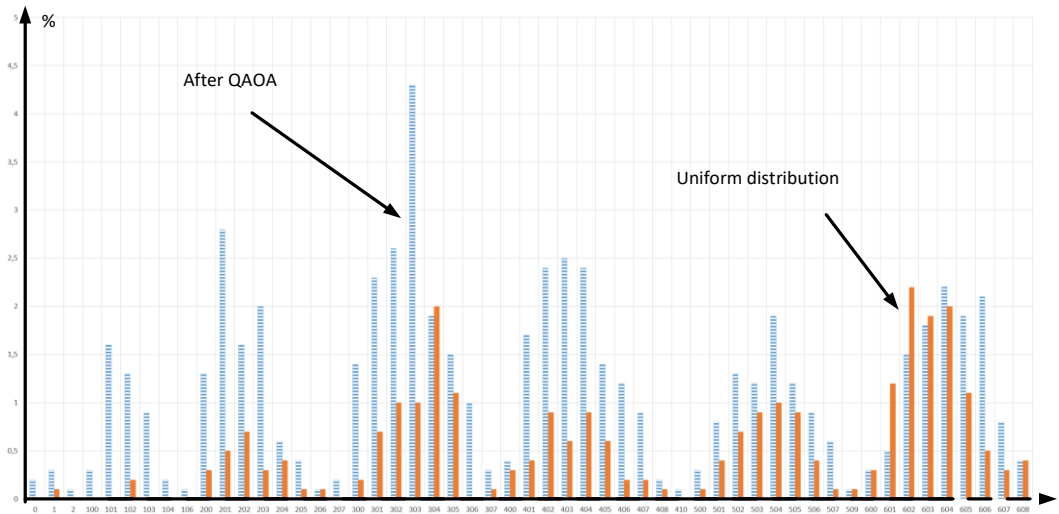

Fig. 5. Distribution of Cost Function values

A careful analysis of the table 2 shows that about 0.6% of the distribution is aggregated to proper or improper solution without penalty cost (> 100) that is about 60 times bigger than the previous value 0.098% which is the value of a uniform state distribution.

The fact that the optimal solutions (solutions with cost 0) have a lowest probability than solutions with cost 1 or 2, is basically linked to the limited number of optimal solutions.

The representation of figure 6 shows how the probabilities of each cost vary between the uniform distribution (in red) used at the beginning of QAOA and the distribution found at the end of QAOA (blue). Higher probabilities are assigned to lower costs at the end of QAOA and lower probabilities are assigned to higher cost proving that QAOA has be successful in controlling the distribution over iterations to high quality states.

The modifications of the distribution shapes of figure 5 can be analyzed considering the ratio between the previous curves (after/before) display in figure 6. This ratio is higher on small cost (that are related to high quality solutions) that reassures that the theoretical based affirmation of QAOA promises meet our experiments.

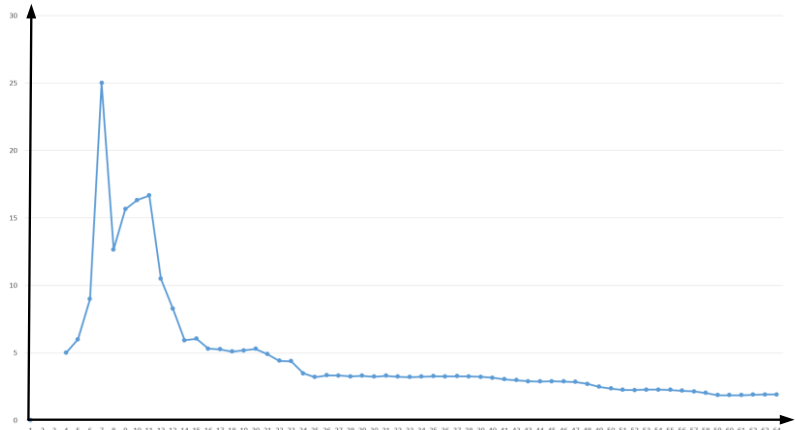

Fig. 6. Overall deformation of probabilities

## 5.2. Numerical experiments with 8 nodes graph

We consider a graph with 8 nodes and 11 arcs introduced in figure 7. The qubit-string space grows to 4 294 967 296 with 4 colors since length are 32 qubits long (maximum size available on the IBM Qiskit simulator). Note that the qubit-strings search space is much larger than the number of graph coloring solutions which does not exceed $4^8 = 65\,536$ for the 8 nodes/4 colors instances (0.0015% of the qubit string search space).

To put it in more concrete terms, there exists a probability of 0.0015% to have a qubit-string in the form 1000 0100 …0001 0010 where only one color is assigned to each node. With 8 colors the qubit-string space is about 167 777 216 strings and the quantum circuit requires only 24 qubits. The number of graph coloring solutions is $3^8 = 6\,561$ for the 8 nodes/3 colors i.e. 0.00391% of the qubit-string search space.

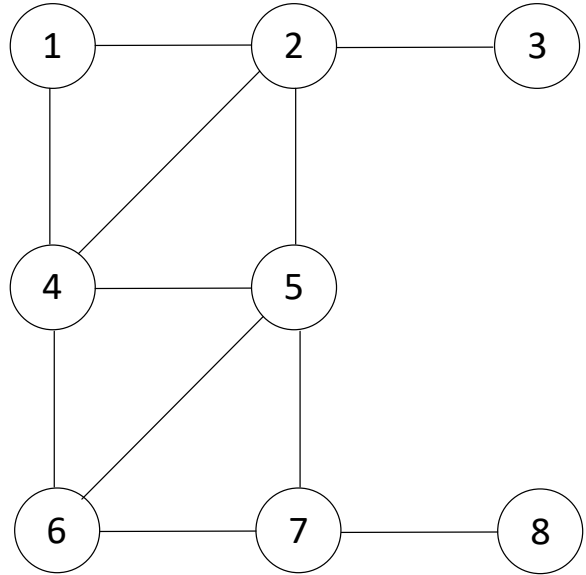

Fig. 7. A graph with 8 nodes, 11 edges to color with 4 colors

For the problem with 3 colors, the following set of parameters has been used:
- $p = 3$ (iterations of Hamiltonian and mixer Hamiltonian);
- $A = 100$ is the weight assigned to the quadratic term in the objective function;
- Cobyla is the iterative numerical method used to minimize the estimated expectation value with 500 iterations;
- 100 samplings is used to estimate $C^p(\vec{\beta^i}, \vec{\gamma^i})$ at step $i$;
- 1000 samplings of $|\psi(\vec{\beta^*}, \vec{\gamma^*})\rangle$ are achieved at the end of QAOA to compute $|s^*\rangle$

**Table 3**
One possible sampling $|\psi(\vec{\beta^*}, \vec{\gamma^*})\rangle$ with 8 nodes, 11 edges.

| Solution cost | Probability 3 colors | Probability 4 colors |
|---|---|---|
| 0 | 0.0 | 0.0 |
| 1 | 0.0 | 0.0 |
| 2 | 0.3 | 0.0 |
| 3 | 0.0 | 0.0 |
| 4 | 0.1 | 0.0 |
| 5 | 0.1 | 0.0 |
| 6 | 0.0 | 0.0 |
| 7 | 0.0 | 0.0 |
| 8 | 0.0 | 0.0 |
| 9 | 0.0 | 0.0 |
| 101 | 0.2 | 0.0 |
| 102 | 0.3 | 0.1 |
| 103 | 0.3 | 0.1 |
| 104 | 0.7 | 0.0 |
| 105 | 0.9 | 0.0 |
| … | … | |

We can note in table 3 that the very low value of $p$ and the small number of samplings affect the QAOA capacity in aggregating the distribution on the optimal and quasi-optimal solutions when the problem scale increases.

Resolution of the graph coloring problem with the 8-nodes graph and 4 colors has a modelization based on 32 qubits which is the larger circuit that can be simulated on the IBM Qiskit simulator. The huge memory consumption in the exponentiation of the operators leads to a large time consuming simulation. One hour has been experienced for the 4 colors instance, on an Intel(R) Xeon(R) at 3.40GHz with 128GO of memory.

### *4.4. Numerical experiments of the "Driven Hamiltonian": 5 nodes graph*

The previous remarks push us into considering that the Driven Hamiltonian should afforded restrictions of exploration to coloring with one color only per node, so that the promise to switch over coloring only that satisfy the constraint C1, enabling us to control which qubit-strings have to be iteratively investigated.

QAOA based on the Driven Hamiltonian has been used with the same set of parameters of section 4.2 and the results of table 4 show that 100% of the distribution is now aggregated on the solution space with one color per node meeting the theoretical considerations. The comparison of QAOA with a Mixer Hamiltonian and a classical Hamiltonian proves the competitive advantages of the Mixer Hamiltonian v.s. the classical one and establishes a numerical validation of the theoretical considerations on the mapping functions.

**Table 4**
Sampling of sampling of $|\psi(\vec{\beta^*},\vec{\gamma^*})\rangle$ with the two Hamiltonians

| Solution cost | Probabilities with the classical mixer operator $p=3$ | Mixer Hamiltonian $p=3$ |
|---|---:|---:|
| 0 | 0.2 | 4.8 |
| 1 | 0.3 | 26.9 |
| 2 | 0.1 | 34.6 |
| 3 | 0.0 | 22.6 |
| 4 | 0.0 | 6.8 |
| 5 | 0.0 | 2.8 |
| 6 | 0.0 | 1.4 |
| 100 | 0.3 | 0.0 |
| 101 | 1.6 | 0.0 |
| 102 | 1.3 | 0.0 |
| 103 | 0.9 | 0.0 |
| 104 | 0.2 | 0.0 |
| 106 | 0.1 | 0.0 |
| … | … | … |

Driven Hamiltonian performances meet the classical well know results in Operation Research community where significant improvements in results have been obtained by both dedicated local search that restrict neighborhood to feasible solutions and by metaheuristic based approaches where the search space investigation is driven by specific information on the solution characteristics.

For example, the split algorithm uses such a trick by transforming one giant trip into a VRP solution allowing next definition of specific moves in the neighbors. Note that the Bierwith's vector (Bierwith, 1996) for the shop-shop is the corner store of the best ever published methods because it permits to investigate only the set of acyclic disjunctive graphs that model solution only avoiding exploration of all disjunctive graph. The best ever published method on the Job-Shop (Nowicki and Smutnicki 1996) takes advantages of this kind of exploration.

The differences between both distributions is strongly significant considering that the probability of a solution of cost 0 is 20 times higher with the Driven Hamiltonian (table 5).

The differences between the two distributions that are apparent in table 4 illustrate that the Mixer Hamiltonian has a substantial advantage by providing a strongly better aggregation on the optimal and quasi-optimal solutions as highlighted on figure 8.

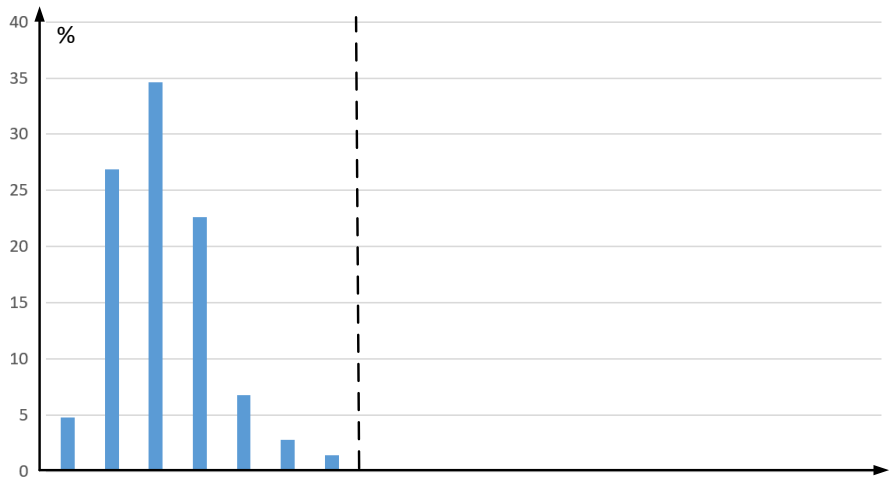

Fig. 8. Comparison of the two probabilities distribution with $p = 3$ for both Hamiltonians

The Driven Hamiltonian offers competitive advantages since it possible to use lower value of $p$ as stressed in table 5.

### Table 5
Sampling of sampling of $|\psi(\vec{\beta^*}, \vec{\gamma^*})\rangle$ with the two Hamiltonians

| Solution cost | Probabilities with the classical mixer operator $p = 3$ | Probabilities with the Driven Hamiltonian $p = 2$ |
|---|---|---|
| 0 | 4.8 | 3.2 |
| 1 | 26.9 | 26.7 |
| 2 | 34.6 | 37.1 |
| 3 | 22.6 | 21.8 |
| 4 | 6.8 | 5.4 |
| 5 | 2.8 | 3.0 |
| 6 | 1.4 | 2.6 |

The global trend of probabilities remains very similar to the trend obtained with $p = 3$ and there is no significant differences between the two probabilities distributions (figure 9) pushing into considering that more compact circuit (low depth circuit) could be used with Driven Hamiltonian.

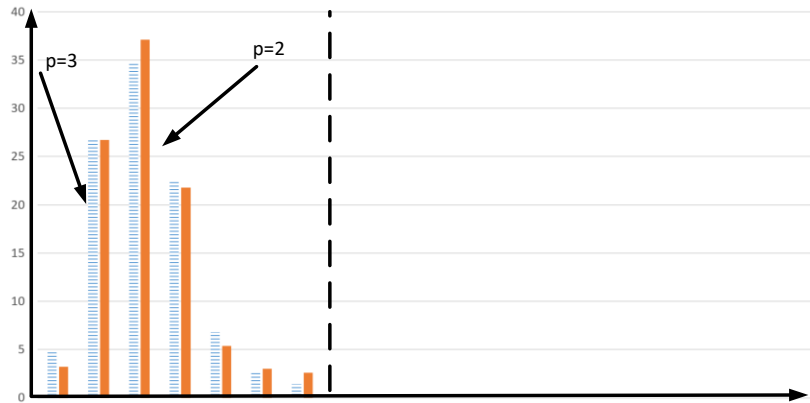

Fig. 9. Comparison of the two probabilities distribution with $p = 2$ for the Driven Hamiltonian

The previous remark must be put in the context of optimization of strongly large scale instances where quantum algorithms make sense. The Driven Hamiltonian enormously simplifies the optimal solution investigation, reducing it to a simple and well-defined feasible search-subspace investigation where a strong deformation of probabilities has been applied to a very short number of strongly high quality solutions.

## 5. Concluding remarks

This work addresses the question of how to implement Mixer Hamiltonians to restrict investigation on a specific subspace for the well-known graph coloring problem that is one of the corner-stone of the operation researcher community. We provided general considerations on Mixer Hamiltonians that are mapping functions defining an indirect representation which is a quantum equivalent of the well-known indirect representation scheme widely used in OR field. The numerical experiments bring into considering the large applicability and efficiency such approaches. To conclude we should emphasis that theoretical considerations and the capability in numerical experiment should push the researchers of the OR community to investigate this new field but keeping in mind that the current quantum computers are currently limited in the number of qubits.